\documentclass[
aip,
jcp,%
amsmath,amssymb,
reprint,%
]{revtex4-1}

\usepackage{graphicx}
\usepackage{bm}

\usepackage{tabu} 

\usepackage{color}
\usepackage{ulem}

\begin{document}

\title{Building Maps in Collective Variable Space}

\author{Ilaria Gimondi}
\affiliation{Thomas Young Centre and Department of Chemical Engineering, University College London.}
\author{Gareth A. Tribello}
\affiliation{Atomistic Simulation Centre, School of Mathematics and Physics,Queen's University Belfast.}%
\author{Matteo Salvalaglio}%
\email{m.salvalaglio@ucl.ac.uk}
\affiliation{Thomas Young Centre and Department of Chemical Engineering, University College London.}%
\date{\today}

\begin{abstract}
Enhanced sampling techniques such as umbrella sampling and metadynamics are now routinely used to provide information on how the thermodynamic potential, or free energy, depends on a small number of collective variables.  The free energy surfaces that one extracts by using these techniques provide a simplified or coarse-grained representation of the configurational ensemble. In this work we discuss how auxiliary variables can be mapped in collective variable (CV) space and how the dependence of the average value of a function of the atomic coordinates on the value of a small number of CVs can thus be visualised.  We show that these maps allow one to analyse both the physics of the molecular system under investigation and the quality of the reduced representation of the system that is encoded in a set of CVs. We apply this approach to analyse the degeneracy of CVs and to compute entropy and enthalpy surfaces in CV space both for conformational transitions in alanine dipeptide and for phase transitions in carbon dioxide molecular crystals under pressure. 
\end{abstract}

\maketitle

\section{Introduction}
Representing the configurational ensemble of a molecular system on a low dimensional hyper-surface defined by a set of collective variables (CVs) $\bm{s}$ is common practice in molecular dynamics simulations. The CVs used to construct these hyper-surfaces can be inspired by physical or chemical intuition, or they might emerge by using a dimensionality reduction algorithm to analyze the trajectory \cite{wales2003energy,das2006low,rohrdanz2011determination,ceriotti2011simplifying}. 
Regardless of how it is constructed, however, the representation in CV space finds its natural application in the analysis of the inherently high dimensional conformational spaces obtained from molecular dynamics trajectories for complex biomolecular systems \cite{das2006low,zhuravlev2009deconstructing,stamati2010application} and for collective transformations in liquids \cite{steinhardt1983bond,tribello2017analyzing,gobbo2017nucleation,Giberti2015,TroutMolOrder,pietrucci2017,pietrucci2011graph}. Moreover, the definition of a set of CVs, $\bm{s}$, as the domain for the definition of bias potentials is commonplace in a range of enhanced sampling methods such as Umbrella Sampling\cite{torrie1977nonphysical,marsili2006self}, Metadynamics\cite{laio2002escaping,WellTempered}, and Adaptive Biasing Force\cite{darve2008adaptive}, to name just a few. 

When CVs are employed to define biasing forces for enhanced sampling methods, the dimensionality of $\bm{s}$ is only limited by the computational efficiency of the sampling protocol.  Furthermore, methods for facilitating the usage of high dimensional sets of CVs have been proposed \cite{piana2007bias,valsson2014variational,pfaendtner2015efficient}.  There is a problem with using a large number of CVs, however, as when analyzing, representing and interpreting information obtained from sampling conformational spaces we are really limited to three dimensions. For this reason, being able to systematically and quantitatively map information on human readable CV spaces is key when it comes to understanding and conveying information on molecular systems. 

In this paper we therefore discuss a set of best practices for mapping auxiliary variables in CV space. This approach allows one to perform a quantitative breakdown of free energy maps into their entropic and enthalpic components, and allows one to map state functions and structural variables along transition pathways. Building such maps in CV space also allows one to assess the local level of degeneracy of the low dimensional representation with respect to auxiliary variables and to thus identify regions where the descriptive quality of the map CVs deteriorates.

The analysis techniques we present are general, and we begin by demonstrating them on a simple 2D model potential. We then assess the accuracy of the method by analyzing the thermodynamics of certain conformational transitions of alanine dipeptide in vacuum. We then conclude the paper by characterizing the I-III polymorphic transition in CO$_2$ molecular crystals under pressure. 

\section{Theory}
\label{sec:conditional}
\subsection*{Mapping variables in CV space with conditional probability}
In what follows the representation of the configuration ensemble of a molecular system on a low-dimensional set of CVs $\bm{s}$ is considered. Additional information for the system mapped on $\bm{s}$ is conveyed by the auxiliary variable $\bar{s}$. To be clear, however, $\bm{s}$ and $\bar{s}$ are both simply functions of the system coordinates. For the sake of clarity in the discussions that follow we will refer to $\bm{s}$ as the map variables and to $\bar{s}$ as the auxiliary variable. 

The equilibrium probability density in the extended domain including both the sampling and auxiliary variables, $p(\bm{s},\bar{s})$, is related to the thermodynamic potential of the ensemble of interest through the Kirkwood relationship\cite{Kirkwood1935}: 
\begin{equation}
F(\bm{s},\bar{s})=-\beta^{-1}\ln{\left(p(\bm{s},\bar{s})\right)}+C
\label{eq:Kirkwood}
\end{equation}
In the canonical ensemble $F(\bm{s},\bar{s})$ is the Helmholtz free energy hyper-surface mapped onto a set of CVs that includes both the map and auxiliary variables. It should be noted, however, that the considerations that follow are general and can be straightforwardly applied to other thermodynamic potentials and their corresponding ensembles, as discussed in the following sections. 

In the domain defined by $\bm{s}$ each point represents an ensemble of configurations that, despite being degenerate in $\bm{s}$, may or may not be identical. For any value of $\bm{s}$ one can therefore define a local probability density for $\bar{s}$.  This local probability density will give one the conditional probability density for $\bar{s}$ subject to a constraint on the value of $\bm{s}$. 

\begin{equation}
p(\bar{s}|\bm{s})={Z_{\bar{s}|\bm{s}}}^{-1} \int e^{-\beta{F(\bm{s},\bar{s})}}\delta(\bm{s}-\bm{s}^\prime) d\mathbf{s}^\prime
\end{equation}
where $Z_{\bar{s}|\bm{s}}$ is a partition function that is locally defined in $\bm{s}$ as: 
\begin{equation}
Z_{\bar{s}|\bm{s}}=\iint {e^{-\beta{F(\bm{s},\bar{s})}}\delta(\bm{s}-\bm{s}{'})d\bar{s}}d\mathbf{s}
\label{eq:local_part}
\end{equation}

This definition of a probability density $p(\bar{s}|\bm{s})$ allows one to systematically map characteristic features of the local distribution of the auxiliary variable $\bar{s}$ onto the domain defined by the map variables, $\bm{s}$.

The most intuitive map that can be constructed provides information on the average value of the auxiliary variable $\bar{s}$. In what follows we indicate this map using the symbol $\langle{\bar{s}}\rangle_{\bm{s}}$ and compute it as: 
\begin{equation}
\langle{\bar{s}}\rangle_{\bm{s}}=\int \bar{s}\,p({\bar{s}|\bm{s}})\,d\bar{s}
\label{eq:average}
\end{equation}
The quantity $\langle{\bar{s}}\rangle_{\bm{s}}$ can be physically interpreted as the ensemble average of $\bar{s}$ computed over the ensemble of configurations that are degenerate in $\bm{s}$. 

One can also compute a map in $\bm{s}$ of any function of the probability density $p({\bar{s}|\bm{s}})$.  For example, one could construct a map for the standard deviation for this distribution as follows:
\begin{equation}
\sigma^{\bar{s}}_{\bm{s}}=\sqrt{\int {\left({\bar{s}-\langle{\bar{s}}\rangle_{\bm{s}}}\right)}^2p({\bar{s}|\bm{s}})  d\bar{s}}
\label{eq:std}
\end{equation}
In the following we will discuss how the appropriate choice of auxiliary variables allows one to analyze transition pathways, to assess the quality of the representation variables and to break down free energy surfaces into their entropic and enthalpic components. 

\begin{figure*}[ht]
\includegraphics[width=0.9\linewidth]{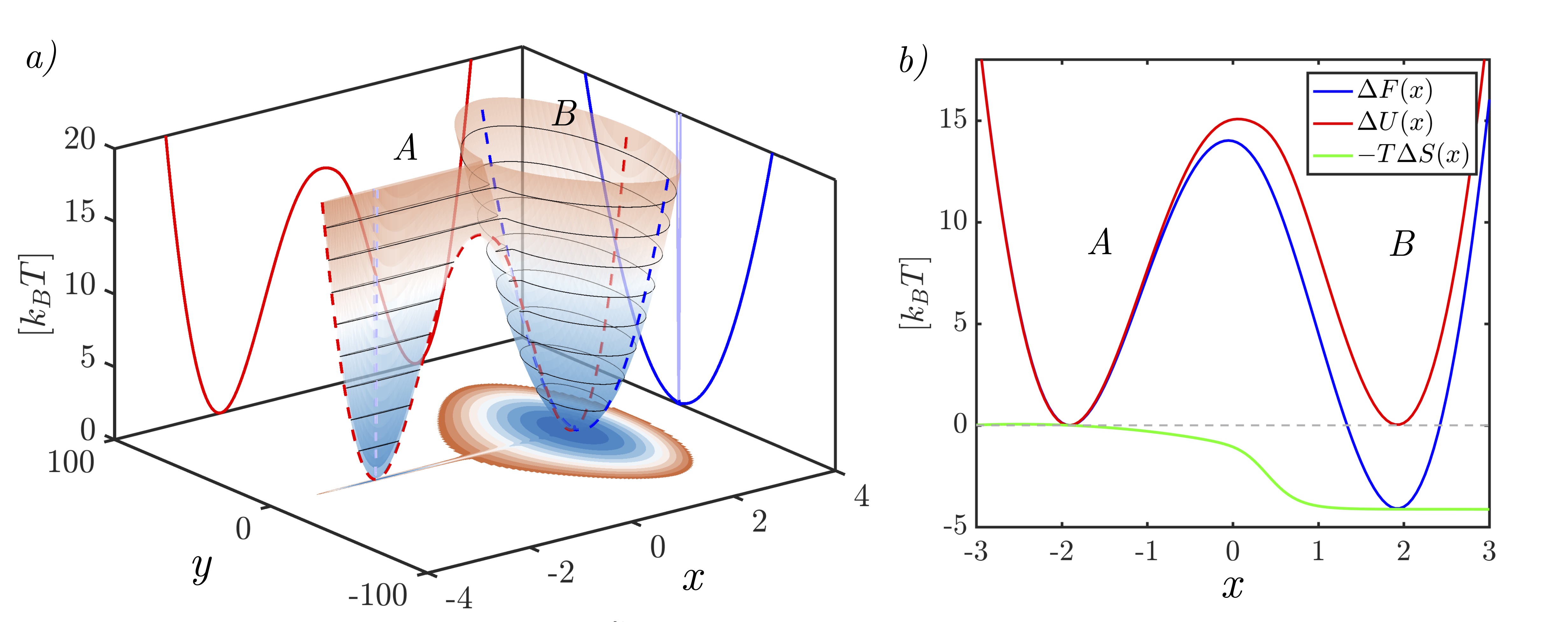}
	\centering
	\caption{(a) 2D model potential function that we have used to explain the technique introduced in this paper.  In what follows we use $x$ as the map variable and $y$ as the auxiliary variable.  As you can see this potential has two minima along $x$ with very similar depths.  Minima B has a much larger width in the auxiliary, $y$, direction than minima A, however. The 2D potential energy function is defined as $E^P(x,y)=-(W_1+W_2)+4x.^2+5\times{10^{-3}}y.^2$, where $W_i=A_i\exp\left({-\frac{(x-x_i).^2}{(2\sigma_i.^2)}-\frac{(y-y_i).^2}{(2\lambda_i.^2)}}\right)$, and $A_1=50.1$, $\sigma_1=1.3$, $\lambda_1=200$, $x_1$=2.5, $y_1$=0, $A_2=50.0$, $\sigma_2=1.3$, $\lambda_2=1.0$, $x_2$=-2.5, and $y_2$=0.(b) Free energy, potential energy and entropy profiles for the potential energy landscape shown in (a) as a function of the map variable $x$.}
	\label{fig:2Dpotential}
\end{figure*}

\subsection*{Mapping Entropy and Enthalpy in CV space}
\label{sec:Entropy}
\paragraph*{Canonical Ensemble.}
A thermodynamic potential surface projected on a CV space $\bm{s}$ implicitly includes both internal energy and entropy contributions.
Often times, however, in order to improve the understanding of molecular processes, to analyze transition pathways in CV space and to infer mechanistic hypotheses, it is desirable to break $F(\bm{s})$ down into its enthalpic and entropic components, and to map each of these separate contribibutions on $\bm{s}$.
To understand how one might go about performing this operation we begin by discussing the decomposition of a thermodynamic potential surface in the canonical ensemble.  In other words, we will discuss how a Helmholtz free energy surface can be decomposed into its entropic and enthalpic contributions.
Consider a scenario in which the exact Helmholtz free energy surface in the map CV domain, $F(\bm{s})=-\beta^{-1}\ln{p(\bm{s})}+C$, has been obtained from the equilibrium probability density $p(\bm{s})$ in the canonical ensemble.
We can write the following expression for this Helmholtz free energy surface, using the definition of the Helmholtz free energy from classical thermodynamics: 
\begin{equation}
F(\bm{s})=U(\bm{s})-TS(\bm{s})
\label{eq:thermodef_helm}
\end{equation}
The term $U(\bm{s})$ here is the internal energy of the ensemble of configurations mapped on $\bm{s}$, $T$ is the temperature of the system, and $S(\bm{s})$ is the entropy of the ensemble of configurations mapped on $\bm{s}$. 
The term $U(\bm{s})$ can be computed as: 
\begin{equation}
U(\bm{s})=\langle{E^P}\rangle_{\bm{s}}+\langle{E^K}\rangle_{\bm{s}}
\label{eq:poten}
\end{equation}
where $\langle{E^P}\rangle_{\bm{s}}$ and $\langle{E^K}\rangle_{\bm{s}}$ are the ensemble averages of the potential and kinetic energies of configurations that are degenerate in $\bm{s}$.  

At this point it is worth noting that $F(\bm{s})$ defines the free energy modulo an immaterial constant $C$. In other words it captures relative free energy differences between ensembles of configurations projected at points of the CV space where $p(\bm{s})$ has been sampled. In order to consistently get rid of $C$ in this work we introduce a reference state $s_{\textrm{ref}}$, which corresponds to an arbitrary point in CV space where $p(\bm{s})$ has been sampled. We thus indicate the relative free energy difference with respect to state $s_{\textrm{ref}}$ as: 
\begin{equation}
\Delta{F(\bm{s})}=F(\bm{s})-F({s_{\textrm{ref}}})=-\beta^{-1}\ln{\frac{p(\bm{s})}{p(s_{\textrm{ref}})}}
\label{eq:ref}
\end{equation}
Incidentally, introducing a reference state ${s}_{\textrm{ref}}$ also allows us to eliminate the $\bm{s}$-independent kinetic energy contribution from Eq. \ref{eq:poten}. It should be noted that this elimination is only possible when $\bm{s}$ only depends on the atomic positions.  That is to say $\bm{s}$ is not a function of the momenta .   If $\bm{s}$ satisfies this condition we can rewrite the internal energy term as: 
\begin{equation}
\Delta{U(\bm{s})}=U(\bm{s})-U({s_{ref}})=\langle{E^P}\rangle_{\bm{s}}-\langle{E^P}\rangle_{s_{\textrm{ref}}}
\label{eq:DU}
\end{equation}
Where $\langle{E^P}\rangle_{\bm{s}}$ is the local ensemble average of the potential energy mapped on $\bm{s}$, which can be directly computed using Eq. \ref{eq:average} with $\bar{s}=E^P$: 
\begin{equation}
\langle{E^P}\rangle_{\bm{s}}=\int_\Omega{E^P}\,p(E^P|\bm{s})\,dE^P
\label{eq:Epot}
\end{equation}
The ensemble average of the potential energy for the reference state that is indicated using  $\langle{E^P}\rangle_{s_{\textrm{ref}}}$, in the expressions above is simply $\langle{E^P}\rangle_{\bm{s}}$ evaluated for $\bm{s}=s_{\textrm{ref}}$.  

Combining Eq. \ref{eq:thermodef_helm}, \ref{eq:ref}, and \ref{eq:DU}, provides an expression that we can use to calculate a map of entropy differences in $\bm{s}$: 
\begin{equation}
\Delta{S}(\bm{s})=\frac{1}{T}\left(\Delta{U}(\bm{s})-\Delta{F}(\bm{s})\right)
\label{eq:entropy}
\end{equation}

\paragraph*{Isothermal-Isobaric Ensemble} 
\label{sec:Enthalpy}
Now consider sampling in the isothermal isobaric ensemble.  Sampling in this ensemble yields a Gibbs free energy map $G(\bm{s})=-\beta^{-1}\ln{p(\bm{s})}+C$, which is defined as:
\begin{equation}
G(\bm{s})=U(\bm{s})-TS(\bm{s})+PV(\bm{s})
\label{eq:thermodef}
\end{equation}

Following the approach that was detailed for the canonical ensemble, we define a common reference state and express the internal energy term using Eq. \ref{eq:DU}.  At variance with the previous section, however, we also introduce a pressure-volume work term that is computed using: 
\begin{equation}
P\Delta{V}\left(\bm{s}\right)=P\left(\langle{V}\rangle_{\bm{s}}-\langle{V}\rangle_{s_{\textrm{ref}}}\right)
\end{equation}

The map in CV space for the local ensemble average of the system's volume $\langle{V}\rangle_{\bm{s}}$ is obviously computed by using Eq. \ref{eq:average} with $\bar{s}=V$: 
\begin{equation}
\langle{V}\rangle_{\bm{s}}=\int_\Omega{V}\,p(V|\bm{s})\,dV
\end{equation}
The enthalpic contribution to the free energy can therefore be mapped as: 
\begin{equation}
\Delta{H}(\bm{s})=\Delta{U}(\bm{s})+P\Delta{V}\left(\bm{s}\right)
\end{equation}
while the map of the entropic contribution to the Gibbs free energy surface defined in $\bm{s}$ is given by:
\begin{equation}
\Delta{S}(\bm{s})=\frac{1}{T}\left(\Delta{H}(\bm{s})-\Delta{F}(\bm{s})\right)
\end{equation}

\subsection*{Entropy map for a 2D model potential}
\label{sec:2Dmodel}
In order to give an intuitive explanation for the physical meaning of the entropy and energy maps that have been introduced in the previous section we will begin by considering a two dimensional model potential.  The functional form, $E^P(x,y)$, for the potential energy landscape that we have studied is given in the caption to Fig \ref{fig:2Dpotential}.  Furthermore, the left panel of the figure gives an illustration of the potential. For the purposes of this example we will use $x$ as the map collective variable.  
In other words, the $x$ variable will be used in a way that is analogous to the way the $\bm{s}$ variable were used in the general discussions of the previous section.  The $y$ variable, by contrast, will be a hidden, unknown variable and will thus be ignored in the analysis of the free energy landscape. 
The reason for using the two variables in this way is that the potential energy landscape $E^P(x,y)$ has two wells whose centers have different values for the $x$ variable and the same value for the $y$ variable.  Figure \ref{fig:2Dpotential}a shows that these two wells, A and B, have width{s} in $x$, that are comparable.  Furthermore, the depths of the two wells are the same.  There is a marked difference between the two wells, however, as the extension in $y$ of well B is markedly larger than that of well A.
It is straightforward to calculate a canonical probability density for this potential in $\mathbf{r}=[x\,y]$ as
\begin{equation}
p(\mathbf{r})=Z^{-1}e^{{-\beta E^P(x,y)}} \quad  \textrm{where} \quad Z=\int{e^{{-\beta E^P(x,y)}}\textrm{d}x\textrm{d}y}.
\end{equation}
We can thus calculate the free energy as a function of the map variable $x$ straightforwardly by using 
\begin{equation}
F(x) = -\beta^{-1}\ln \left[ \int e^{-\beta E^P(x,y)} \textrm{d}y\right].
\end{equation}
The resulting free energy profile that we obtain by applying this equation is shown in blue in the right panel of figure \ref{fig:2Dpotential}.  When the free energy landscape is projected in this way basin B appears to have a free energy that is substantially lower than the free energy of basin A, which is perhaps surprising given that we know that potential energies of these two basins are the same.  We can understand the physical origin of this effect, however, by constructing the average potential energy and entropy maps that were discussed in the previous section.  The red line in figure \ref{fig:2Dpotential}b is the potential energy profile along $x$ which was computed using Eq. \ref{eq:average},  and the potential energy as an auxiliary variable i.e. $\bar{s}=E^P$, while the green line is the entropy map.  

Elementary statistical mechanics tells us that the entropic term is large when the accessible volume of phase space is larger.  The results described in the previous paragraph were thus to be expected.  In fact the same conclusion could have been drawn by simply examining the energy landscape and noting that basin B has a larger spatial extent in the auxiliary variable $y$.  It is obvious that these greater spatial extents are going to ensure that the degeneracy of $x$ with respect to the hidden variable $y$ is larger in basin B than it is in basin A and that the entropy of basin B is thus going to be larger than the entropy of basin A.  The example is still instructive, however, because, while these considerations are intuitively obvious for a a model potential with a single hidden degree of freedom, they are far less obvious in real systems, where degeneracy with respect to hidden variables can often be very difficult to quantify.  In these cases Eq.\ref{eq:entropy} is thus very useful as it allows one to map this hidden entropic contribution in CV space and to thus disentangle the roles played by entropy and energy in ensuring thermodynamic stability. 

\section{Simulation Details}
\label{sec:SimulationDetails}
In order to demonstrate our analysis we discuss its application, in different flavors, to two {model systems; namely,}  Alanine dipeptide in vacuum and the I-III polymorphic transition of solid CO$_2$ under pressure.  In analysing these systems  we have to deal with the fact that we do not have an exact, analytical expression for the free energy surface and thus have to extract this quantity via sampling.  This limitation introduces two practical problems which we will discuss how to resolve.  The first of these problems is that we will only have a finite number of samples.  It will thus be important for us to quantify the random error in all our estimates of the ensemble averages.  The second practical problem is that many of the conformational transitions that we are interested in take place over timescales that are far longer than we can simulate using molecular dynamics.  In what follows, we will thus, after a brief discussion of the simulations that we have performed, discuss how we can use bias potentials to enhance the rates for these slow processes and how the free energy can be extracted from such biased simulations.

\begin{figure}[ht]
	\includegraphics[width=1\linewidth]{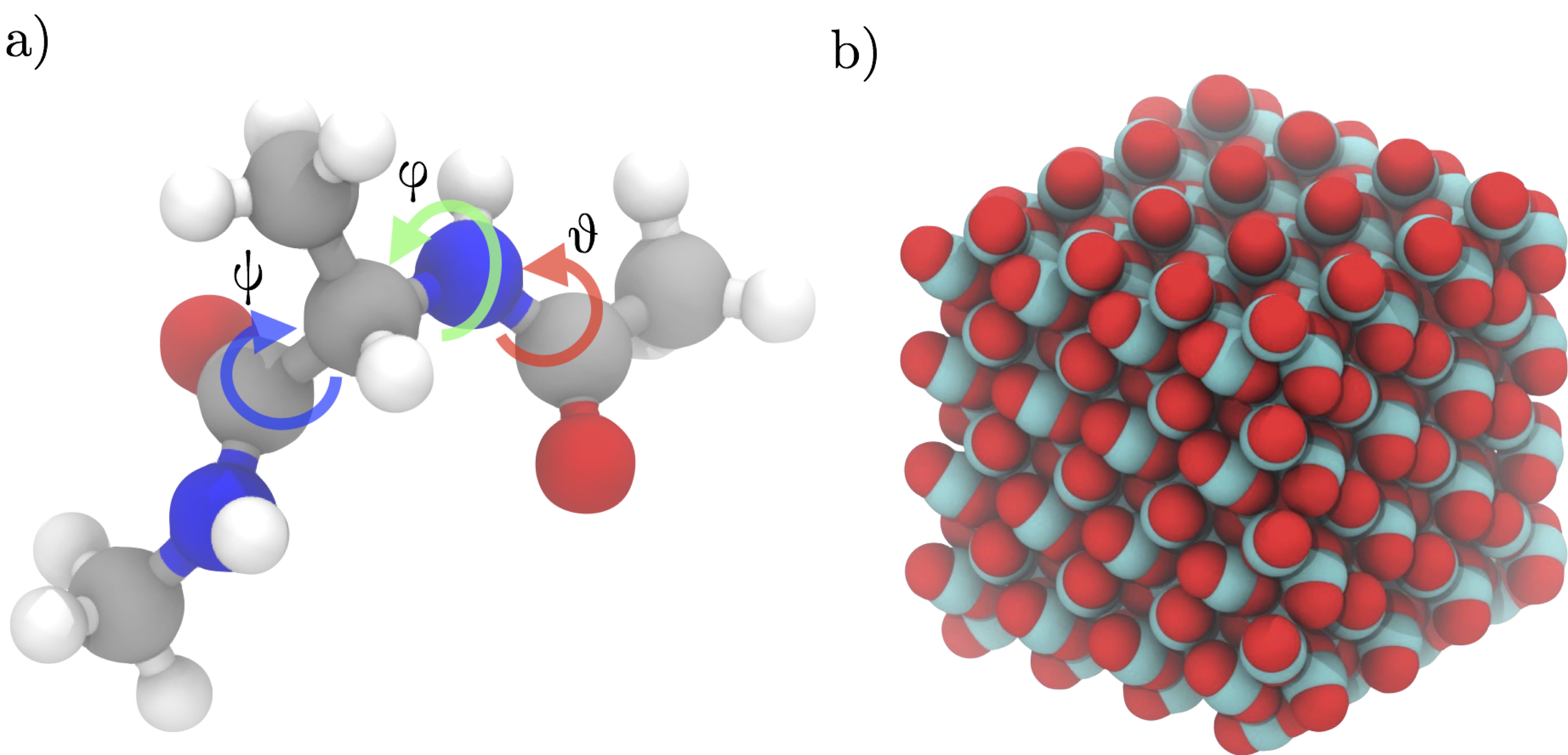}
	\centering
	\caption{a) {A representation of the} Alanine Dipeptide molecule that has been studied in this work that illustrates the three dihedral angles $\phi$, $\psi$, and $\theta$ that are instrumental in our analysis. b) A snapshot showing Phase I of crystalline CO$_2$.  This particular phase is the thermodynamically stable state at 3 GPa and 350K.}
	\label{fig:molecules}
\end{figure}

\subsection*{Alanine dipeptide in vacuum}
Four independent well tempered metadynamics (WTmetaD) \cite{WellTempered} simulations of alanine dipeptide in vacuum have been carried out at T=300, 350, 400, and 450K. The AMBER99SB \cite{Wang2004} force field was used in all these calculations, periodic boundary conditions were not applied, and non bonded interactions were computed with an infinite cutoff. All bonds were constrained using LINCS, and the dynamics was propagated using a time-step of 2~fs for 2$\mu$s for each temperature. Temperature was controlled with the Bussi-Donadio-Parrinello thermostat, and initial velocities were randomly chosen from the Maxwell-Boltzmann distribution at the appropriate temperature\cite{Bussi2007}. 
WTmetaD was carried out by depositing Gaussians in a CV space defined by the dihedral angles $\phi$ and $\psi$ (see Fig. \ref{fig:Aladipeptide1}) every 500 steps.  These Gaussians had an initial height of 1.2~kJ~mol$^{-1}$ and a width of 0.35 rad for both CVs.  The bias factor for well-tempered metadynamics was set equal to 6 and all simulations were performed with gromacs 5.1.4 \cite{gromacs} patched with plumed 2.3 \cite{plumed2}

\subsection*{CO$_2$ polymorphic transitions}
The I - III polymorphic transition in solid carbon dioxide was investigated using Well-Tempered Metadynamics at 350 K {and} 3 GPa. The initial configuration was a super cell of phase I that was composed of 64 unit cells (256 molecules).  The unit cell of phase I was obtained from the Crystallography Open Database (COD, ID 1010060), from ref \cite{Smedt1924}, while for phase III it was built from refs \cite{Kuchta1988,Etters1989}. This initial structure was minimized and then a 500 ps NVT equilibration was performed at 350 K. This initial equilibration was followed by a 5 ns NPT equilibration at 350 K and 3 GPa without long-range corrections and a 5 ps NPT equilibration with long-range corrections. 
The TraPPE force field\cite{POTOFF1999} was employed which necessitated the introduction of two dummy atoms per molecule in order to ensure that each molecule remained rigid with the desired 180$^\circ$ angle.\cite{Sanghi2012} Periodic boundary conditions were employed, the cut-off was set to 0.7 nm and long-range corrections were included for both the Van-der-Waals and the electrostatics by using the \textit{particle mesh Ewald} (\textit{pme}). To ensure that the isothermal and isobaric ensemble is simulated the Bussi-Donadio-Parrinello thermostat\cite{Bussi2007} was used together with an anisotropic Berendsen barostat.\cite{Berendsen1984} The integration timestep was set equal to 0.5 fs, and WTmetaD was run for $\sim$281 ns.

In the well-Tempered Metadynamics simulations the bias was deposited on two collective variables. These CVs are the order parameters that are based on the local environment around each CO$_2$ molecule that are discussed in \cite{Salvalaglio2012,Giberti2015_2,gimondi2017co2}.  In essence these variables measure whether the coordination numbers and the relative orientation between pairs of molecules in each others first coordination shells are similar to the arrangements that are found in the perfect crystal. A detailed description of the formulation of $\lambda_I$ and $\lambda_{III}$ can be found in our previous work on CO$_2$\cite{gimondi2017co2,gimondi2018co2}. In the metadynamics simulations the well tempered bias factor is set equal to 100 and Gaussians are deposited every 500 steps.  These Gaussians have an initial height {of} 10~kJ~mol$^{-1}$ and a width, $\sigma$, of $7.81 \times 10^{-3}$ in both CVs. 
In addition, a harmonic repulsive potential is used to ensure that the box sides only fluctuate between 1.7 and 3.0 nm.  This potential prevents excessive and irreversible distortions of the cell shape that can occur when the system undergoes a transition to the melt. It is worth mentioning, however, that at the T-P conditions investigated in this work (350 K - 3 GPa) no melting is observed and so the box edges do not approach regions of configuration space where these constraints would act.
When analyzing the CO$_2$ trajectories we often discuss the anisotropy of the supercell, which is defined as the ratio between the lengths of the largest and the smallest edges of the simulation box.

\subsection*{Reweighting methods and Conditional Probability Convergence}

As discussed at the start of section \ref{sec:SimulationDetails} the timescales for many of the processes that we are interested in are often longer than we can simulate.  As discussed in the methodology section above we therefore use metadynamics to enhance the frequencies with which these rare events occur in our simulations.  Our usage of this technique introduces a history-dependent bias potential on the map variables, $V(\mathbf{s},t)$.  Consequently, if we assume that the the system is equilibrium with the bias potential at all times the probability, $p'(\mathbf{s},t)$, that we sample a particular set of map variables at any given point in our trajectory is given by:
\begin{equation}
p'(\mathbf{s},t) = \frac{e^{-\beta F(\mathbf{s}) -\beta V(\mathbf{s},t)} }{\int e^{-\beta F(\mathbf{s}) -\beta V(\mathbf{s},t)} \textrm{d}\mathbf{s} }
\label{eqn:biased-p}
\end{equation}
It is straightforward to show \cite{souaille2001extension,shirts2008statistically,tan2012theory} that, if the bias' time dependence is ignored and if the final, converged metadynamics bias, $V(\mathbf{s})$, is used in place of $V(\mathbf{s},t)$ in the above, an estimate of the unbiased free energy can be extracted from a set of $M$ trajectory frames using:

\begin{equation}
F(\mathbf{s}) = -\beta^{-1} \ln \left[ \frac{\sum_{i=1}^M \delta(\mathbf{s}-\mathbf{s}_i) e^{+\beta V(\mathbf{s}_i) } }{\sum_{i=1}^M e^{+\beta V(\mathbf{s}_i) } } \right]
\label{eqn:reweight}
\end{equation}

Recently a number of more refined techniques \cite{tiwary2014time,REWE} that take the time dependence of the bias into account have been proposed.  The aim of this section is, therefore, to compare the performance of these different methods.  These new methods for reweighting start from a  recognition that equation \ref{eqn:biased-p} can be written as:

\begin{equation}
p'(s,t) = \frac{e^{-\beta F(\mathbf{s}) - \beta V(\mathbf{s},t) - \beta c(t) }}{\int e^{-\beta F(\mathbf{s})} d\mathbf{s}}
\end{equation}
with
\begin{equation}
\quad c(t) = \beta^{-1} \ln\left[ \frac{\int e^{-\beta F(\mathbf{s})} d\mathbf{s}}{\int e^{-\beta F(\mathbf{s}) - \beta V(\mathbf{s},t)} d\mathbf{s}} \right]
\end{equation}

and thus introduce $c(t)$ as a running estimate of the difference between between the normalisation constants for the unbiased and biased probability distributions.  As the bias potential is time dependent this quantity is obviously also time dependent. It is, therefore, useful to estimate its time dependence in order to make best use of the statistics that were collected before the bias had fully converged.  
In Tiwary and Parrinello's method \cite{tiwary2014time} this is achieved by updating the estimate of $c(t)$ every $\tau$~ps of the trajectory using:

\begin{equation}
c(t) = \beta^{-1} \ln \left[ \frac{\int e^{ \gamma \Delta \beta V(\mathbf{s},t)} d\mathbf{s}}{ \int e^{\Delta \beta V(\mathbf{s},t)} d\mathbf{s} } \right]
\end{equation}

where $\gamma$ is the well-tempered metadynamics parameter and where $(\Delta \beta)^{-1}=k_B T (\gamma - 1)$.  In  Tiwary and Parrinello's method reweighting is thus achieved by using equation \ref{eqn:reweight} with $V(\mathbf{s})=V(\mathbf{s},t) + c(t)$.  Bonomi et al. \cite{REWE} use a different method to deal with $c(t)$ and introduce the following expression that describes how the  probability that a CV, $\mathbf{f}$, takes a particular value over a time window changes over a time period of length $\Delta t$:

\begin{equation}
P(\mathbf{f},t + \Delta t) = \int e^{-\beta\Delta t\left[ V'(\mathbf{s},t) - \overline{V'}(t) \right]} P(\mathbf{s},\mathbf{f},t) d\mathbf{s}
\end{equation}

In this expression $V'(\mathbf{s},t)$ is the derivative of the bias potential with respect to time at the map variable $\mathbf{s}$ and $\overline{V'(t)}$ is an average of this time derivative that is calculated by integrating over the whole domain.  In the method of Bonomi et al. a histogram that is a function of $f$ and $\mathbf{s}$ is therefore accumulated and a suitably-manipulated version of the expression above is used to convert the biased histogram that is accumulated back to the unbiased distribution.  In the software that was released with Bonomi's paper the numerical details of this procedure can be done in one of two, notionally-equivalent, ways so we test both in what follows.   

Before constructing maps for any of the systems that were simulated using metadynamics we performed a short study on our alanine dipeptide data to compare the efficacy of these various reweighting algorithms.
In this section we will thus analyze the data from a WTmetaD simulation of alanine dipeptide in vacuum at 300 K.  In particular, the free energy estimates that we obtained by applying the various reweighting methods described in the previous paragraph were compared against the free energy estimate obtained by integrating the time dependent WTmetaD bias potential.  In what follows we use the symbol $\Delta{F(\phi,\psi)}_{ref}$ to refer to the estimate of the free energy that was obtained by integrating in this way.

\begin{figure*}[ht]
	\includegraphics[width=0.75\linewidth]{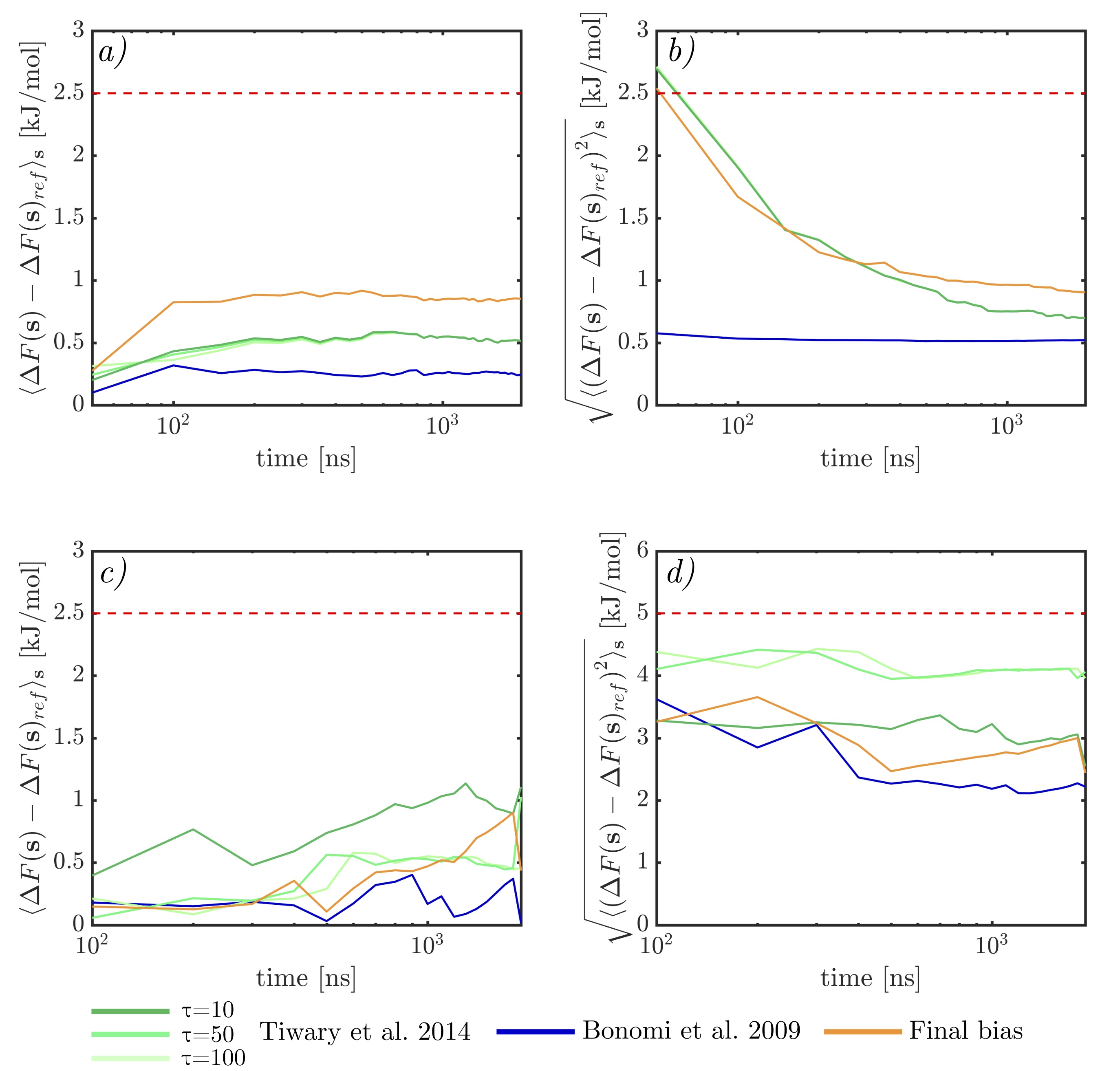}
	\centering
	\caption{Convergence of conditional probability distributions in CV space for alanine dipeptide in vacuum at T = 300 K.  To construct the figures in the top two panels we estimated the free energy as a function of the map variables, $\bm{s}$, using three different reweighting techniques. We then computed the $\bm{s}$-dependent difference with respect to the FES obtained by integrating the bias potential. In a) we report the average value of the difference, in b) we report its standard deviation. It can be seen that in all cases both these indicators are well below $k_BT$, showing that all the reweighting strategies provide consistent estimates of the probability density in $\bm{s}$. In panels (c) and (d) we repeat this procedure but now, instead of computing the free energy as a function of the map variables only, we compute it as a function of the map variables and the auxiliary variable, $E^P$.  The way the reference free energy is constructed in this second case is explained in the text.  Also in this case the mean difference is well below $k_BT$, while the width of the residuals distribution is of the order of $k_BT$.} 
	\label{fig:convergence_rewe}
\end{figure*}

In the analysis of the convergence of the free energy as a function of the Ramachandran angles, $\Delta F(\phi,\psi)$, that is reported in Fig. \ref{fig:convergence_rewe} we consider using the final bias in equation~\ref{eqn:reweight}, the method proposed by Tiwary and Parrinello \cite{tiwary2014time} with three different $\tau$ parameters (10, 50, and 100 metadynamics cycles) and the implementations of Bonomi et al's method that are included in the reweight utility that is distributed with PLUMED 1.3 \cite{PLUMED,plumed2}.  
To assess the degree to which each of these methods has converged to the reference result we define residuals $\delta=\Delta{F(\phi,\psi)}-\Delta{F(\phi,\psi)}_{ref}$.  In the upper panels of figure \ref{fig:convergence_rewe} we monitor how the mean and standard deviation of the residual distribution varies as a function of the simulation time. 

In the lower two panels of figure \ref{fig:convergence_rewe} the degree to which $\Delta{F(\phi,\psi,E^P)}$ has converged is monitored.  A slightly different approach was used when constructing these two figures as the free energy as a function of $\phi$, $\psi$ and $E^P$ cannot be calculated by integrating the simulation bias because $E^P$ was not biased in the simulation. In this case we thus used the average of the estimates obtained by employing the final bias, the approach of Bonomi et al.\cite{REWE}, and the approach of Tiwary and Parrinello\cite{tiwary2014time} on the entire 2 $\mu{s}$ trajectory as a reference FES.  Once again, panels (c) and (d) show the mean and the standard deviation for the distribution of residues.

Fig. \ref{fig:convergence_rewe}a shows that the average absolute discrepancy between the reweighted FES and the FES that is obtained by integrating the deposited bias is within $k_BT$.  
Furthermore, this discrepancy always converges to a fraction of {a}~kJ~mol$^{-1}$ within the first 50 ns of the simulation. It should be noted that the method used to construct histograms has a significant impact on the high energy regions of the FES leading to large deviations from the FES obtained by integrating the metadynamics bias. 
To limit this effect histograms for the Tiwary and Parrinello, and final bias reweighting methods were constructed using Gaussian kernels with a bandwidth of 0.035 rad. 

The comparisons in figure \ref{fig:convergence_rewe} show that when comparing with the FES obtained by integrating the metadynamics bias the reweighting approach of Bonomi is slightly more accurate.  When estimating probability density with respect to auxiliary variables, however, all  the methods tested provide a very similar degree of accuracy. We note that, in practice, even simply reweighting with the final bias gives an accurate estimate of the free energy surface. 

\subsection*{Estimating sampling errors}

At the start of section \ref{sec:SimulationDetails} we explained that, we cannot extract an exact, analytical expression for the free energy surface and that we instead use simulations to estimate this quantity by sampling.  This procedure introduces a sampling error that it is important to quantify. We can estimate this using a block averaging technique \cite{frenk-smit02book}, which divides our trajectory up into a series of $N$ blocks of length $M$. 
If we are calculating a map for the average value of the auxiliary variable, $\overline{s}$ we need to consider the propagation of the sampling error associated with Eq.\ref{eq:average}.  
To this aim we calculate the following two functions of the map variables for each of our blocks of trajectory data:     
\begin{equation}
\left( Z_{\overline{s}|\mathbf{s}} \left\langle \overline{s}(\mathbf{s}) \right\rangle \right)^{(j)} = \frac{1}{W^{(j)}}  \sum_{t=1}^M \overline{s}_t \delta( \mathbf{s}_t - \mathbf{s} ) e^{+\beta V(\mathbf{s},t)}  
\end{equation}
\begin{equation}
Z_{\overline{s}|\mathbf{s}}^{(j)} = \frac{1}{W^{(j)}}  \sum_{t=1}^M \delta( \mathbf{s}_t - \mathbf{s} ) e^{+\beta V(\mathbf{s},t)}
\end{equation}

where $W^j=\sum_{t=1}^M e^{+\beta{V(\mathbf{s},t)}}$

We can obviously recover the averages of either of these two quantities over the whole trajectory from the quantities calculated for each block using:  

\begin{equation}
\langle A \rangle = \frac{\sum_{i=1}^N W^{j} A^{j}}{\sum_{i=1}^N W^{j}}
\end{equation}

where $A^{(j)}$ in the above is substituted by either $\left(w\bar{s}(\mathbf{s})\right)^j$ or $W^j$. 

Better still, however, because we set the blocks lengths to be longer than the autocorrelation time for both $\mathbf{s}$ and $\overline{s}$ we can compute an estimate for the variance using: 

\begin{equation}
\delta^2(A) = \frac{\Omega}{\Omega - S/\Omega} \sum_{j=1}^N W^{j} \left( A^{j} - \langle A \rangle \right)^2
\end{equation}

where $\Omega$ and $S$ are the sum and the sum of the squares of the $W^{(j)}$ values for each block. As we have samples from each of our $N$ blocks we can thus compute a confidence limit on our estimate of $\langle A \rangle$ using:

\begin{equation}
\epsilon = \Phi^{-1}\left( \frac{p_c + 1}{2} \right) \sqrt{\frac{\delta^2(A)}{N}}
\end{equation}

where $\Phi^{-1}$ is the inverse of the cumulative probability distribution for a standard normal distribution and where $p_c$ is the level of statistical confidence we would like our error bars to represent.  This quantity, $\epsilon$, must be estimated for the two averages $Z_{\overline{s}|\mathbf{s}} \left\langle \overline{s}(\mathbf{s}) \right\rangle$ and $Z_{\overline{s}|\mathbf{s}}$ separately.  The error on the map $\langle \overline{s}(\mathbf{s} )\rangle$ is then given by:

\begin{equation}
\epsilon\left( \overline{s}(\mathbf{s} ) \right) = \langle \overline{s}(\mathbf{s} )\rangle \sqrt{ \frac{\epsilon(Z_{\overline{s}|\mathbf{s}} \left\langle \overline{s}(\mathbf{s}) \right\rangle)}{Z_{\overline{s}|\mathbf{s}}{ \left\langle \overline{s}(\mathbf{s}) \right\rangle} }  + \frac{\epsilon(Z_{\overline{s}|\mathbf{s}})}{Z_{\overline{s}|\mathbf{s}}} }
\end{equation}

\begin{figure*}[ht]
	\includegraphics[width=1.0\linewidth]{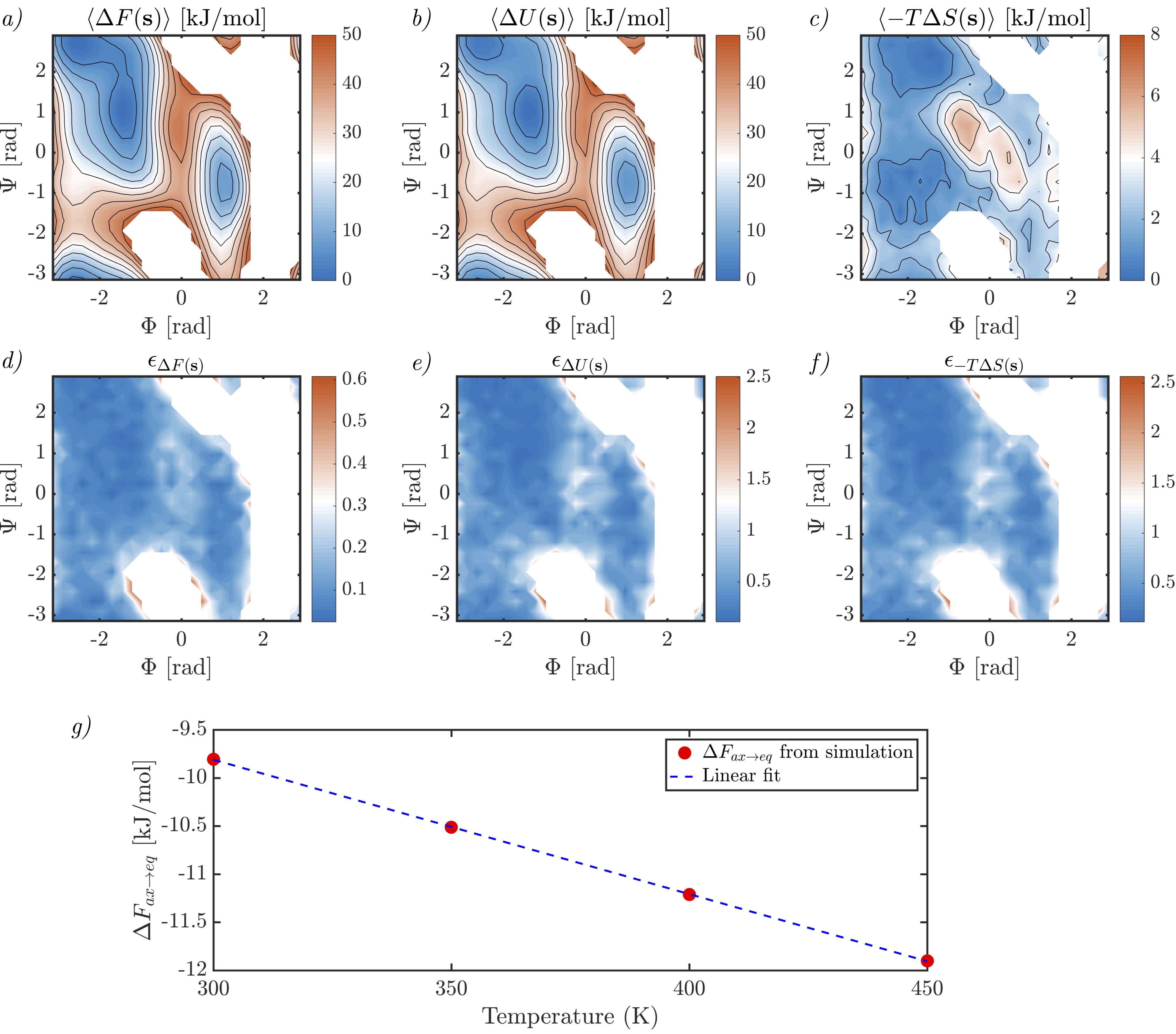}
	\centering
	\caption{Breakdown of the free energy surface for alanine dipeptide in vacuum at T=300~K. We denote block averaged maps using angular brackets. Block averaged $\Delta{F{(\bm{s})}}$ (a), $\Delta{U{(\bm{s})}}$ (b) and $-T\Delta{S{(\bm{s})}}$ (c) are reported as a function of the Ramachandran CV space $\bm{s}=(\phi ,\psi)$, together with maps of their respective sampling errors $\epsilon_{\Delta{F{(\bm{s})}}}$ (d), $\epsilon_{\Delta{U{(\bm{s})}}}$ (e), and $\epsilon_{-T\Delta{S{(\bm{s})}}}$ (f). The sampling errors, averaged over the CV space for configurations which have $\Delta{F}<50$ kJ/mol, are  $\overline{\epsilon_{\Delta{F}}}=0.14$, $\overline{\epsilon_{\Delta{U}}}=0.61$, and $\overline{\epsilon_{T\Delta{S}}}=0.63$  kJ/mol.
	 (c) Free energy difference between the $C7_{ax}$ and $C7_{eq}$ conformations as a function of temperature.  The red circles correspond to estimates obtained from simulations performed at four different temperatures.  The blue dashed line is the result of fitting these four data points using a linear function.  The parameters $\Delta{U}$ and $T\Delta{S}$ obtained from this fit are reported in table \ref{tab:DeltaS} and compared with estimates obtained from a single simulation using Eq.\ref{eq:Epot} and \ref{eq:entropy} .}	\label{fig:Aladipeptide1}
\end{figure*}

\section{Results}

\section*{Alanine Dipeptide in vacuum}
Alanine dipeptide in vacuum provides a prototypical example of a conformational free energy landscape that is characterized by metastable states. As such dialanine has often been used as a case study for development and as a test bed for enhanced sampling algorithms.  In this section we will thus use this system once more in order to discuss a few aspects of our analysis in a setting that should be familiar to other researchers in this field.  In addition, by simulating this simple system we should be able to assess the quantitative accuracy of the ensemble averages mapped in CV space that we will obtain. 
We will begin by comparing the entropy and internal energy differences between the metastable states of alanine dipeptide (C7$_{eq}$ and C7$_{ax}$).  This analysis is useful as the internal energy and entropy differences between these two states can be computed in one of two ways.  We can compute the internal energy and entropy maps described in section \ref{sec:Entropy} from a single simulation at one particular temperature and hence extract the internal energy and entropy difference between the two states.  Alternatively, we can calculate free energy surfaces at a range of temperatures and extract from these the difference in free energy between the two states as a function of temperature.  Equation \ref{eq:entropy} tells us that this free energy difference should be a linear function of temperature and that the intercept and gradient of this line should be equal to the internal energy difference and the entropy difference respectively.  In what follows we will thus compute the energy and entropy differences in these two ways in order to test the reliability of our approach. Furthermore, we also show that we can use our approach to accurately compute the FES at a temperature that is different from the one we simulated at. Finally, we discuss how projecting auxiliary variables that describe the conformation of the molecule allows one to further analyze the complexity of the ensemble of configurations projected in CV space. 

\begin{figure*}[ht]
	\includegraphics[width=1.0\linewidth]{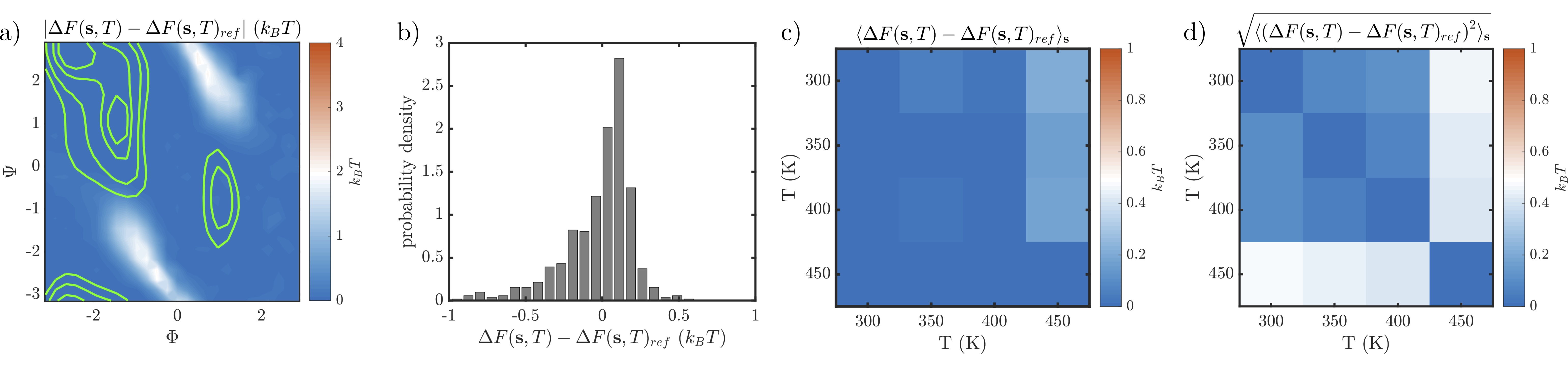}
	\centering
	\caption{A further test on the efficacy of our new method.  To construct the first two of these figures the free energy surface $\Delta{F(\bm{s},T)}$ was computed at T=450 K by applying Eq. \ref{eq:thermodef} with $\Delta{S}(\bm{s})$ and $\Delta{U}(\bm{s})$ values that were computed from a simulation that was performed at 350 K.  In addition, a reference free energy surface  $\Delta{F(\bm{s},T)_{ref}}$ was computed by analyzing a WTmetaD that was performed at T = 450 K. Panel (a) shows the difference between these two free energy surfaces as a function of the $\phi$ and $\psi$ variables.  Panel (b) then shows the cumulative distribution of the local differences. Panels (c) and (d) show what happens when this analysis is extended over all the temperatures studied.  The colors of the squares in panel (c) are used to indicate the average difference between the estimate of the free energy that is obtained at the temperature shown on the vertical axis by using Eq. \ref{eq:thermodef} with $\Delta{S}(\bm{s})$ and $\Delta{U}(\bm{s})$ surfaces that are computed at the temperature on the horizontal axis with an estimate of the free energy that is computed from a simulation at the temperature shown on the vertical axis.  Panel (d), meanwhile, shows a similar set of results but in this panel the standard deviation for this distribution of differences is shown instead of the mean.  It is clear from these figures that the average differences between these various estimates are all significantly lower than $k_B T$ at all temperatures.}
	\label{fig:Aladipeptide2} 
\end{figure*}

\subsection*{Conformational transition thermodynamics}
To assess the reliability of the free energy breakdown into its enthalpy and entropy components that is obtained by mapping the potential energy as described in section \ref{sec:Entropy} we analyzed a set of four independent WTmetaD simulations that were carried out at 300, 350, 400, and 450 K. 
For each simulation we computed the change in internal energy $\Delta{U}_{ax\rightarrow{eq}}$ and {the change in} entropy $\Delta{S}_{ax\rightarrow{eq}}$ {that is} associated with the $C7_{ax}\rightarrow{C7_{eq}}$ conformational transition.  The values that we obtained by performing these analyses were then compared with estimate{s} {for} $\Delta{U}_{ax\rightarrow{eq}}$ and $\Delta{S}_{ax\rightarrow{eq}}$ that were obtained by fitting the  dependence of $\Delta{F}_{ax\rightarrow{eq}}$ on temperature using a linear function.  This linear fit is justified because we know that the free energy is given by \ref{eq:thermodef_helm}.  Furthermore, Fig. \ref{fig:Aladipeptide1} shows that the data points obtained from the four simulations all lie very close to the linear regression line.  Table \ref{tab:DeltaS} shows that the parameters we obtained from this fit are consistent with the four independent estimates for $\Delta{U}_{ax\rightarrow{eq}}$ and $\Delta{S}_{ax\rightarrow{eq}}$ that were obtained by applying the approach detailed in Section \ref{sec:Entropy} to the simulations at the four different temperatures.  To obtain values for the internal energy differences, $\Delta{U}_{ax\rightarrow{eq}}$, in this table  we computed the difference between the ensemble averages of $\Delta{U}(\phi,\psi)$ that were computed over the domains in CV space corresponding to conformer $C7_{eq}$ and conformer $C7_{ax}$, respectively. The $\Delta{S}_{ax\rightarrow{eq}}$ term was instead computed using Eq.\ref{eq:entropy}. 

\begin{figure*}
\includegraphics[width=1.0\linewidth]{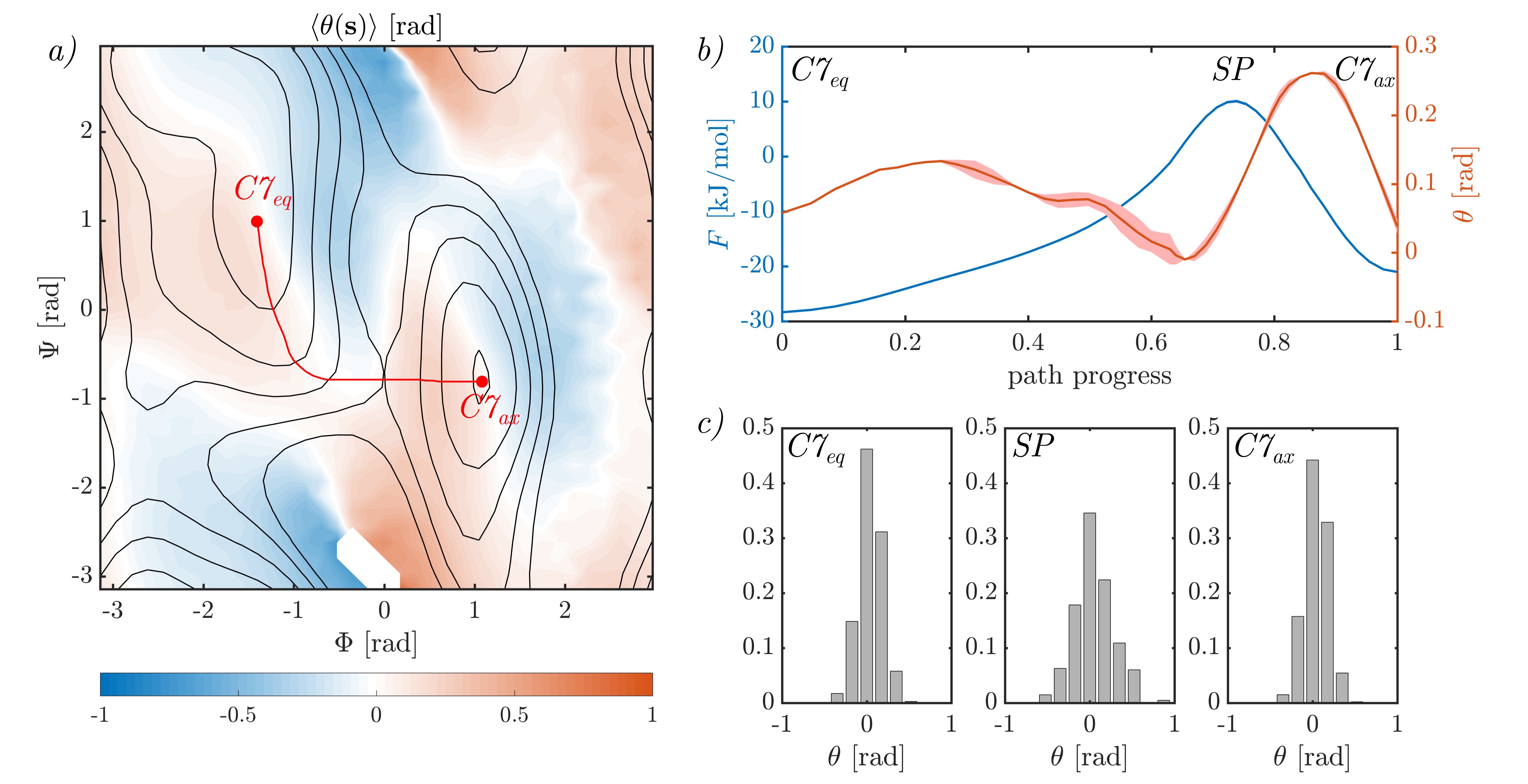}
	\centering
	\caption{a). Map {showing} the {average value of the} $\theta$ dihedral angle {as a function of the} dihedral angles $\phi$, and $\psi$. The average sampling error computed on the map is $\overline{{\epsilon}_\theta}=0.0256$ rad. b) Value of the $\theta$ dihedral angle mapped along the pathway between $C7_{eq}$ and $C7_{ax}$ represented in red in panel a).  In constructing this figure the sampling error was computed at the 95\% confidence intervals and is reported as a shaded area. c) Probability distributions for $\theta$ at specific points in $\phi$,$\psi$ space.  {The particular points chosen} correspond to the two (meta)stable conformers and {to} the  saddle point $SP$ corresponding to the apparent transition state in CV space.}
	\label{fig:Aladipeptide3}
\end{figure*}

\begin{table}
\caption{Internal energy and entropy changes associated with the $C7_{ax}\rightarrow{C7_{eq}}$ conformational transition for alanine dipeptide in vacuum.  Values reported in the first column in this table were obtained by fitting multiple simulations at different T while others were obtained by applying equations \ref{eq:Epot}, and \ref{eq:entropy} to the results from single simulations.}
\begin{small}
\begin{tabular}{cc|cccc}
\hline

 & $\Delta{F}_{{ax}\rightarrow{{eq}}}(T)$& &  Eq. \ref{eq:Epot}, \ref{eq:entropy} \\ 
 & {linear fit} & 300 K & 350 K & 400 K & 450 K \\
\hline
 $\Delta{U}_{{ax}\rightarrow{{eq}}}$ & -5.6219 &  -5.5619 &   -5.5662 &   -5.7285 &   -6.0741      \\
 $\Delta{S}_{{ax}\rightarrow{{eq}}}$ & -0.0140 & -0.0142 &   -0.0141 &    -0.0137 &    -0.0129    \\
 \hline 

\end{tabular} 
\end{small}
\label{tab:DeltaS}
\end{table}

\subsection*{Computing temperature-dependent Free Energy surfaces}

If a FES can be broken down into its internal energy $\Delta{U}(\bm{s})$ and entropy $\Delta{S}(\bm{s})$ components as shown in Fig. \ref{fig:Aladipeptide1}a-c, it becomes trivial to capture the local temperature dependence of the free energy projected in CV space. In other words, if one has the internal energy and entropy maps at one temperature it is straightforward to compute the free energy surfaces $\Delta F(s)$ at a second, different temperature by locally applying Eq. \ref{eq:thermodef}\cite{Laio2009}. As a second test for this methodology we thus performed a comparison between free energy surfaces that were obtained using Eq. \ref{eq:thermodef} with those obtained by re-simulating the system at the various different temperatures.  In particular, we computed $\Delta{F}(\phi,\psi)_{i}$, $\Delta{U}(\phi,\psi)_i$ and $\Delta{S}(\phi,\psi)_i$ for each of the four WTmetaD simulations performed at T$_i$=300, 350, 400, and 450 K. 
From each of the pairs of maps that we obtained from this analysis we then computed
 $\Delta{F}(\phi,\psi)_{j,i}=\Delta{U}(\phi,\psi)_i-T_j\Delta{S}(\phi,\psi)_i$.  Fig. \ref{fig:Aladipeptide2} shows the results of a comparison between the free energy surfaces that are obtained through this procedure and the reference free energies that are computed by simulating at each of the temperatures, $\Delta{F}(\phi,\psi)_{i}$.  The first panel of this figure shows a typical map of the absolute difference between $\Delta{F}(\phi,\psi)_{j,i}$ and $\Delta{F}(\phi,\psi)_{i}$ with i=450 K and j=350 K.  It is clear from this figure that there is quantitative agreement between the free energy surfaces that are computed in these two different ways. The residual differences between the two estimates are of the order of one tenth of $k_BT$ for most of the map.  In fact the only regions where there are substantial, 2 $k_BT$ discrepancies, between the two free energy estimates are the highest energy regions of the energy landscape.

To further illustrate the quantitative agreement between these two different ways of estimating the free energy, Fig. \ref{fig:Aladipeptide2}b shows an example of the cumulative distribution of the absolute error. It is clear from this figure that for the majority of grid points the discrepancy between the two estimates is less than 0.5 $k_B T$.  Furthermore, figure \ref{fig:Aladipeptide2} also shows that similar results hold for other pairs of temperatures.  To extract figure \ref{fig:Aladipeptide2}c we computed $\Delta{F}(\phi,\psi)_{j,i}$ for each of the 16 possible combinations of temperatures.  For each of these surfaces we then performed a comparison that was similar to that shown in panel (a) between $\Delta{F}(\phi,\psi)_{j,i}$ and $\Delta{F}(\phi,\psi)_{i}$.  This procedure gives us a map showing how the difference between the two estimates of the free energy depends on $\phi$ and $\psi$.  Rather than displaying all this information for all 16 possible pairs of temperatures we calculated the mean and standard deviation of each of the 16 set of local difference values.  In figures \ref{fig:Aladipeptide2}(c) and \ref{fig:Aladipeptide2}(d) four by four grids are used to display the values of these 16 means and 16 standard deviations respectively. {It is clear from these two figures that} both the mean {and the} standard deviation {of the differences} are smaller than 0.5 $k_BT$.  There is thus quantitative agreement between the values obtained using the two methods over a wide range of temperature{s}. We note that the effectiveness of the approach described relies on the fact that all the relevant states are present in the entire temperature range investigated, and hence the number of basins and their location in CV space do not depend on temperature. While somehow restrictive we expect these conditions to hold for most conformational transition problems. 

\subsection*{Mapping Auxiliary Structural Variables}

In addition, to computing maps that show how the entropy and the internal energy depend on the map variables, $\bm{s}$, we can also calculate maps for any auxiliary CV $\bar{s}$. Fig. \ref{fig:Aladipeptide3} shows why this procedure is useful.  In this figure the average value for the $\theta$ dihedral angle of alanine dipeptide is shown as a function of the Ramachandran angles $\phi$ and $\psi$. It is clear from this plot that some details of the interconversion mechanism between the $C7_{eq}$ and $C7_{ax}$ is hidden when the free energy landscape is displayed as a function of the Ramachandran angles, which is interesting given that this energy landscape is considered to be fully understood. Figure \ref{fig:Aladipeptide3} shows that both the $C7_{eq}$ and $C7_{ax}$ basins are characterized by two sub-populations of states {that can take both} positive and negative values for the angle $\theta$.
Furthermore, within both the $C7_{eq}$ and $C7_{ax}$ basins one can see an anti-correlation between the local average of the value of $\theta$ and the value of $\phi$.  This anti-correlation is reminiscent of the anti-correlation between the trajectories that escape from the $C7_{eq}$ basin and those that commit to the $C7_{ax}$ basin\cite{bolhuis2000reaction,Tiwary2013}.

It is also interesting to map the local average of $\theta$ onto the minimum free energy path for the $C7_{eq}\rightarrow{C7_{ax}}$ transition.  The result of performing this calculation is shown in figure \ref{fig:Aladipeptide3}(b).  It is interesting to note that the average value of $\theta$ changes sign along the pathway. Furthermore, the most probable $\theta$ angle for configurations that are projected on top of the free energy barrier in CV space is close to zero. 

In addition, to calculating the average value of the auxiliary structural variable as a function of the map variables we can also compute the conditional probability distribution for the auxiliary variable for a particular set of map variables, $p(\theta|\phi,\psi)$.  Figure \ref{fig:Aladipeptide3}(c) reports three such conditional probability distributions that have $\phi$ and $\psi$ values that correspond to being in the two stable conformers and at the apparent transition state (TS). It is clear from these figures that the width of the distribution of $\theta$ values depends markedly on the position in $(\phi,\psi)$. As a case in point when the system is at the apparent TS the width of the local distribution in $\theta$ is much larger.  This is perhaps because in these conditions the system is between the two basins and will thus undergo larger fluctuations.

\begin{figure*}[ht!]
	\includegraphics[width=0.97\linewidth]{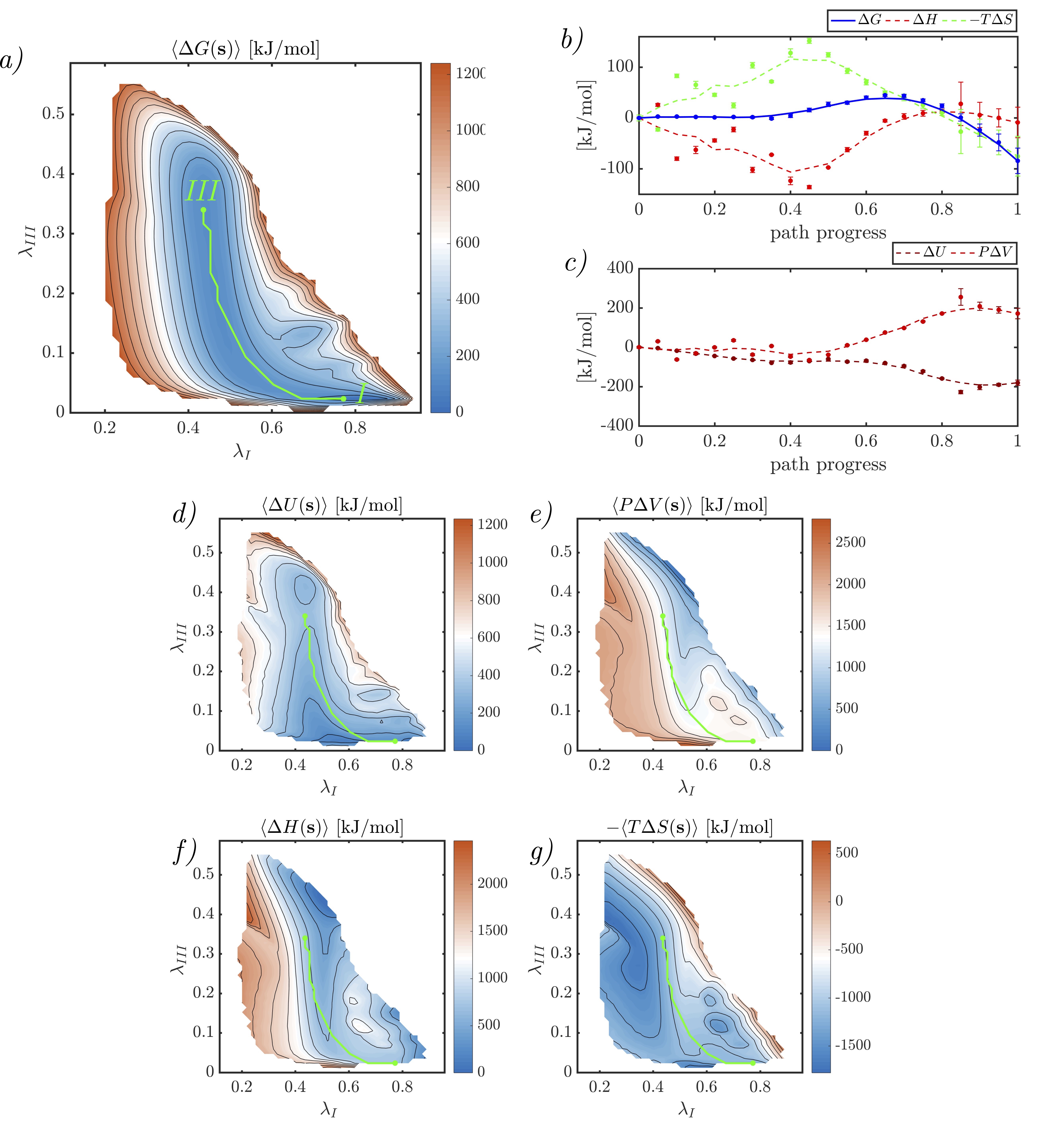}
	\centering
	\caption{Breakdown of the free energy surface (a) associated with the polymorphic transition between form I and III for solid $CO_2$. Block averaged $\Delta{U}$ (d), $P\Delta{V}$ (e), $\Delta{H}$ (f), and $-T\Delta{S}$ are reported as a function of the collective variables $\lambda_I$ and $\lambda_{III}$ that were introduced and discussed at length in Ref. \cite{gimondi2017co2}.	In all the maps a green line indicates the minimum free energy path for the I-III transition. 
The sampling errors averaged over the CV domain for these surfaces are $\overline{\epsilon_{\Delta{G}}}=17.14$, $\overline{\epsilon_{\Delta{U}}}=16.3$, $\overline{\epsilon_{\Delta{H}}}=28.3$ and $\overline{\epsilon_{T\Delta{S}}}=35.07$ kJ/mol.
    f-g) Breakdown of the free energy difference between polymorph I and III of solid $CO_2$ along the minimum free energy path in which the error bars indicate 95\% confidence intervals. In all plots the transition path is defined so that moving rightwards corresponds to moving from form III to form I.  Panel (b) {shows that} form I has a lower free energy at 350 K and 3 GPa {than form III}. Intriguingly, however, the overall enthalpy contribution is close to null and the fact that form I is more stable in these conditions must, therefore, be due to entropic contributions.
Panel (c) shows that the internal energy and the mechanical work of expansion along the $I\rightarrow{III}$ transition pathway almost compensate for each other. While the $P\Delta{V}$ favours form III, internal energy favours form I along the entire transition pathway.}
	\label{fig:CO2maps}
\end{figure*}

\begin{figure*}[ht]
	\includegraphics[width=1\linewidth]{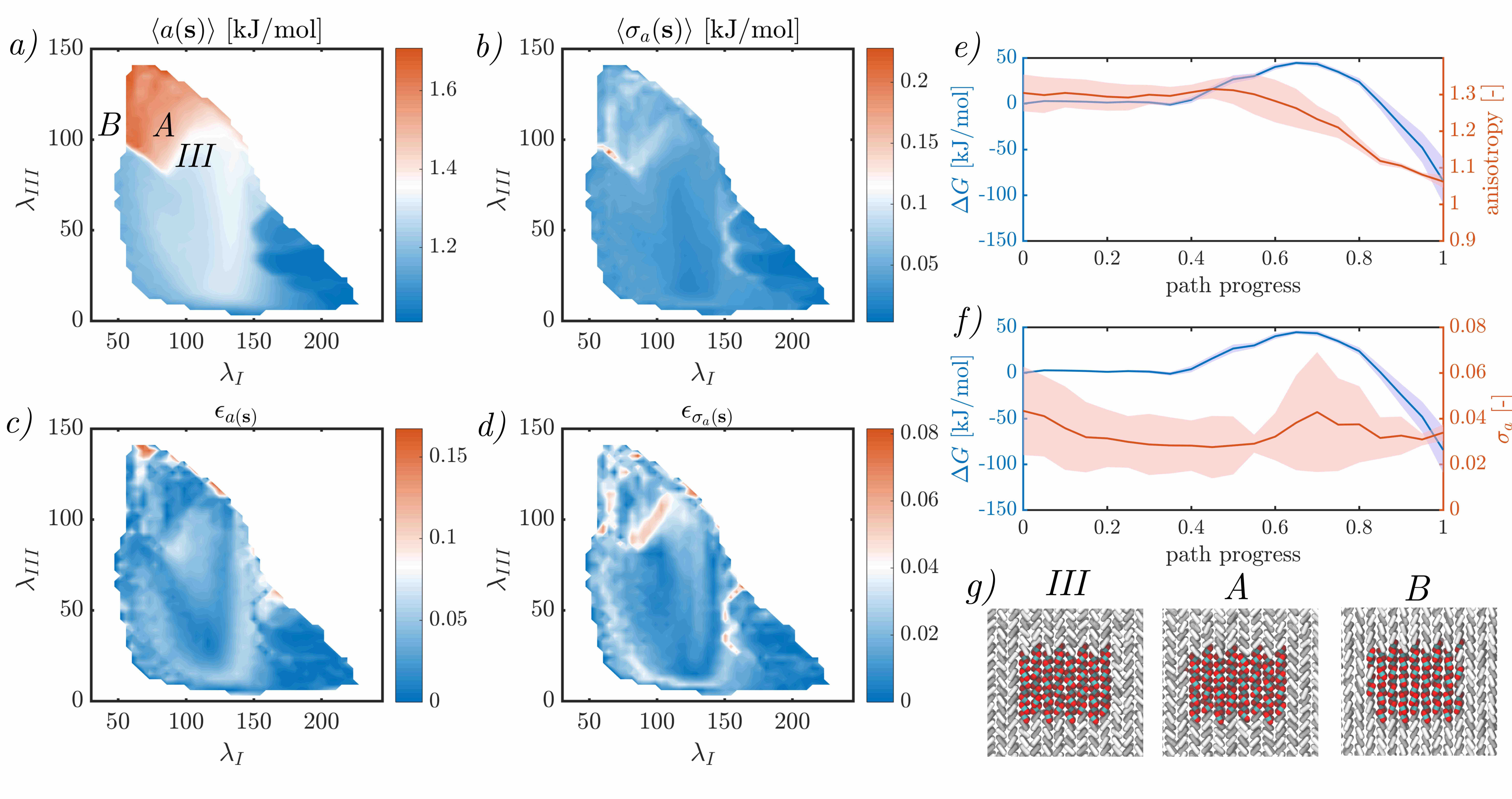}
	\centering
	\caption{Maps of structural variables in CV space: cell anisotropy map (a) and map of the standard deviation of the cell anisotropy (b) with their respective error maps (c,d). e) Cell anisotropy along the minimum energy pathway showing that cell anisotropy is highly correlated with the free energy profile. f) Standard deviation of the cell anisotropy along the minimum energy pathway. In both (e) and (f) sampling error, reported as 95\% confidence interval, is shown as a shaded area. Interestingly the maximum for the fluctuations in the cell anisotropy along the transition pathway occurs at the same point as the top of the free energy barrier. Moreover, these figures show that the sampling error of the fluctuation is maximal near the top of the free energy barrier. g) Typical configurations projected in the CV space in the proximity of phase III. $A$ corresponds to a view of a slightly distorted phase III configuration, while $B$ corresponds to an unstable ordered packing {that is} projected in this region of CV space.}
	\label{fig:anisotropymap}
\end{figure*}

\section*{I-III polymorphic transition in CO$_2$}
\subsection*{Polymorphic transition thermodynamics}
As discussed in section \ref{sec:Enthalpy} we can break down any free energy maps that was obtained by simulating in the isothermal isobaric ensemble by building maps in CV space for the average potential energy and the average volume. In this section we will thus map the local enthalpic and entropic contributions to the Gibbs free energy in CV space and thus investigate the I-III polymorphic transition of solid CO$_2$. In a recent paper \cite{gimondi2017co2} we have investigated the thermodynamics and the mechanism for this phase transition using a rigid model of CO$_2$. Here we further analyze simulations that are performed at conditions for which form I is thermodynamically stable with respect to form III, i.e. 3 GPa and 350K, and will produce maps in the space defined by the collective variables $\lambda_I$ and $\lambda_{III}$ (see section \ref{sec:SimulationDetails}, and ref. \cite{gimondi2017co2} for details). Fig. \ref{fig:CO2maps} shows the free energy $\Delta{G(\lambda)}$, internal energy $\Delta{U(\lambda)}$, mechanical work $P\Delta{V(\lambda)}$, and entropy $-T\Delta{S(\lambda)}$ maps. 
Fig. \ref{fig:CO2maps} shows that all of these terms contribute significantly to the free energy surface.  The features that we observe in the final free energy surface cannot therefore be attributed to one of the particular components from which this quantity is determined.  Furthermore, there is a clear transition channel around the minimum free energy path for all the surfaces shown in figure \ref{fig:CO2maps}.
It is also interesting to note that, {there is a small free energy barrier between form III and form I for values of the progression variable of around 0.6.} {By breaking down this free energy landscape} into {its constituent enthalpic and entropic} components {we can} clearly identify {that this} barrier {is} associated with the enthalpy contribution, and in particular with the mechanical work of expansion $P\Delta{V}$ (see Fig. \ref{fig:CO2maps}c) that is necessary to transform form III into form I. On the other hand, entropic contributions contribute crucially to the  stabilization of form I.

\subsection*{Assessing degeneracy in CV space.}
Thus far, the maps in CV space that we have discussed and analyzed have been based on the calculation of local ensemble averages using Eq. \ref{eq:average}. While ensemble averages for auxiliary structural variables allow us to improve the description of the configuration ensemble that is projected in $\bm{s}$, it is the width of the conditional probability density $p(\bar{s}\vert\bm{s})$ that provides the information on the local level of degeneracy in the CV space. As suggested in section \ref{sec:conditional}, however, any function of $p(\bar{s}\vert\bm{s})$ can be computed and mapped in $\bm{s}$.  One can thus also apply Eq.\ref{eq:std} to $\bm{s}$ and thus map the standard deviation of $p(\bar{s}\vert\bm{s})$. This procedure is useful as the standard deviation does indeed provide information on the local width of the conditional probability density in $p(\bar{s}\vert\bm{s})$. 
Since the I-III polymorphic transition in the bulk takes place through a concerted rearrangement of CO$_2$ molecules that is assisted by a global anisotropic expansion of the crystal super-cell, we decided to consider the system' anisotropy as auxiliary variable to analyse degeneracy in CV space.
We constructed the map of the standard deviation of the cell anisotropy $\sigma_{A}$ in the space defined by the collective variables $\lambda_I$, $\lambda_{III}$ that is reported in Fig. \ref{fig:anisotropymap}b. Moreover, we use{d} this map to reconstruct the behavior of $\sigma_{A}$ along the minimum free energy path that connects phase I to phase III in the $\lambda_I$,$\lambda_{III}$ space (\ref{fig:anisotropymap}b). 
The $\sigma_{A}$ profile along the minimum free energy path {clearly shows that} $p(\bar{s}\vert\bm{s})$ tends to broaden as the system moves away from phase I. The standard deviation $\sigma_{A}$ {then} goes through a local maximum {whose position appears to correspond to the position of the} apparent transition state associated with the transformation to phase III. 
The presence of this local maximum in $\sigma_{A}$ {would appear to} indicate that the degeneracy with respect to the anisotropy of configurations is larger when the system is close to the saddle point.  In other words, the fluctuations in the anisotropy is small when the system is in phase I or phase III and large when it is at the apparent transition state between these two states.
This behaviour suggests that {the} $\lambda_I$ and $\lambda_{III}$ CVs, which were built to distinguish between the unperturbed structures of polymorph I and III, {are less descriptive} in {the} regions {of phase space} where transitions take place.  Furthermore, this may also suggest that the projection of different ensembles of structures overlap in the map variables space. 
This observation is consistent with results that we have recently obtained for this system by performing a committor analysis and histogram tests in $\lambda_{I}$,$\lambda_{III}$ \cite{gimondi2017co2}.  These calculations revealed that, while the $\lambda_{I}$ and $\lambda_{III}$ variables can provide a satisfactory description for the I-III transition thermodynamics, {they} cannot properly map the transition state ensemble. To map the transition state ensemble one must include the system anisotropy explicitly as discussed at length in Ref.\cite{gimondi2017co2}. Our point here is that with the analysis of the anisotropy maps carried out in this section we obtain a qualitatively similar insight at a fraction of the computational cost. 

By further analysing the $\sigma_A$ map one can clearly see additional regions where the standard deviations in this quantity are large and where the map variables are thus perhaps deficient. 
In addition to the apparent TS region, large fluctuations in the cell anisotropy are evident in the region of CV space marked with the label A in Fig. \ref{fig:anisotropymap}a-d.  {By} analyzing the configurations {that are} projected in this region we can see that {these} wider distributions  {for} anisotropy are associated with the projection of distorted phase III configurations (snapshot A in Fig. \ref{fig:anisotropymap}). Furthermore, moving downhill from region A in Fig. \ref{fig:anisotropymap}b towards the point indicated using the label B we can identify high energy, unstable ordered packings that do not resemble phase III and that do not correspond to a local minimum in the free energy.
In this case it would therefore seem that the topology of the $\sigma_A$ surface allows one to identify a transition region between two different ensembles of structures that are projected at different points in the CV space even when such structures do not correspond to stable free energy basins.

\section{Conclusions}

In this work we described an approach to map auxiliary variables in CV space that works by evaluating local conditional probability densities. By carrying out this type of analysis one can considerably deepen the insight obtained from enhanced sampling simulations and enable an in depth analysis of the thermodynamics of the ensembles of molecular structures projected in CV space.  In addition, it also enables one to critically assess the characteristics of the CV space that has been used to represent the results. To demonstrate our techniques we have analyzed a simple 2D potential, alanine dipeptide in vacuum as well as a polymorphic transition that takes place in CO$_2$ at high pressure. Analyzing these model systems has allowed us to demonstrate how this method can be used to construct internal energy and volume maps in CV space and how these maps can be used to systematically breakdown free energy differences in their energetic, enthalpic and entropic components. Furthermore, {we have shown how} having access to entropy and internal energy maps in CV space allows us to compute free energy surfaces at temperatures different from those at which the conformational space has been sampled with quantitative accuracy. 
In taking this analysis further we have demonstrated how we can construct maps based on the values of auxiliary variables in CV space and how we can use such maps to characterize the evolution of state functions and structural features along transition pathways.  This application is useful as it allows us to identify correlations between variables and to identify the dominant driving force for complex processes such as CO$_2$ polymorphic transitions. Finally we highlight that, by complementing maps of local ensemble averages with higher order features of the conditional probability distribution, one can qualitatively assess the quality of the representation of complex conformational spaces in low-dimensional CV spaces. In particular we can clearly identify regions of the CV space in which different ensembles of molecular structures overlap. 

\section*{Acknowledgements}
The authors acknowledge EPSRC (Engineering and Physical Sciences Research Council) for a Ph.D. scholarship and the UCL Legion High Performance Computing Facility for access to Legion@UCL and associated support services that were used to complete this work.  GAT also acknowledges acknowledges funding from EPSRC (Grant EP/P005004/1).

\section*{References}

\bibliography{MAPS}

\end{document}